%% file: main.tex
\newcommand{\extended}[1]{}    %
\newcommand{\short}[1]{#1}     %

\documentclass[runningheads]{llncs}

\usepackage[T1]{fontenc}
\usepackage{graphicx}
\usepackage{booktabs}
\usepackage{lmodern}
\usepackage{times}
\usepackage{soul}
\usepackage{url}
\usepackage[hidelinks]{hyperref}
\usepackage[utf8]{inputenc}
\usepackage[small,skip=5pt]{caption}
\usepackage{multirow}
\usepackage{float}

\usepackage{
	packages/wojtek15logics,
	packages/wojtek15other,
	packages/macros,
	packages/uppaal,
	packages/prooftree,
}

\definecolor{tucgreen}{RGB}{0,140,79}
\newcommand{\YK}[1]{{\color{blue}#1}}
\newcommand{\WJ}[1]{{\color{tucgreen}#1}}

\newcommand{\unwrap}{\mathcal{M}}

\newcommand{\dott}{\ .\ }

\newcommand{\NC}{NC}
\newcommand{\NB}{NB}
\newcommand{\NV}{NV}
\newcommand{\NMO}{NMO}
\newcommand{\NEC}{NEC}

\newcommand{\NCand}{\ensuremath{\mathit{\NC}}}
\newcommand{\NVot}{\ensuremath{\mathit{\NV}}}
\newcommand{\NMun}{\ensuremath{\mathit{\NMO}}}
\newcommand{\NElec}{\ensuremath{\mathit{\NEC}}}

\newcommand{\MO}{MO}
\newcommand{\EC}{EC}

\newcommand{\EPack}{EP\xspace}
\newcommand{\EPacks}{EPs\xspace}
\newcommand{\BEnv}{BEnv\xspace}
\newcommand{\BEnvs}{BEnvs\xspace}
\newcommand{\REnv}{REnv\xspace}
\newcommand{\REnvs}{REnvs\xspace}
\newcommand{\EDay}{EDay\xspace}

\newcommand{\bstuff}{P1} %
\newcommand{\valvote}{P2} %
\newcommand{\moblock}{P3} %
\newcommand{\tallypropname}{tally integrity\xspace} %
\newcommand{\virt}{m} %

\usepackage{adjustbox}
\usepackage{fancyvrb}
\usepackage{float}

\usepackage{dcolumn}
\newcolumntype{d}[1]{D{.}{.}{#1} }

\newcommand{\setdisplayskips}[1][0pt]{%
\setlength{\abovedisplayskip}{#1}%
\setlength{\belowdisplayskip}{#1}%
\setlength{\abovedisplayshortskip}{#1}%
\setlength{\belowdisplayshortskip}{#1}%
}

\begin{document}
\title{Verification of the Socio-Technical Aspects of Voting: \\
       The Case of the Polish Postal Vote 2020}
\titlerunning{Verification of Socio-Technical Aspects of Voting}
\author{%
Yan Kim\inst{1}
\and
Wojciech Jamroga\inst{1,2}
\and
Peter Y.A. Ryan\inst{1}
}
\institute{
	Interdisciplinary Centre for Security, Reliability, and Trust, SnT, University of Luxembourg\and
	Institute of Computer Science, Polish Academy of Sciences, Warsaw, Poland\\
	\email{\{yan.kim, wojciech.jamroga, peter.ryan\}@uni.lu}
}

\maketitle

\begin{abstract}
Voting procedures are designed and implemented by people, for people, and with significant human involvement. Thus, one should take into account the human factors in order to comprehensively analyze properties of an election and detect threats. In particular, it is essential to assess how actions and strategies of the involved agents (voters, municipal office employees, mail clerks) can influence the outcome of other agents' actions as well as the overall outcome of the election.

In this paper, we present our first attempt to capture those aspects in a formal multi-agent model of the Polish presidential election 2020. The election marked the first time when postal vote was universally available in Poland. Unfortunately, the voting scheme was prepared under time pressure and political pressure, and without the involvement of experts. This might have opened up possibilities for various kinds of ballot fraud, in-house coercion, etc. We propose a preliminary scalable model of the procedure in the form of a Multi-Agent Graph, and formalize selected integrity and security properties by formulas of agent logics. Then, we transform the models and formulas so that they can be input to the state-of-art model checker Uppaal. The first series of experiments demonstrates that verification scales rather badly due to the state-space explosion. However, we show that a recently developed technique of user-friendly model reduction by variable abstraction allows us to verify more complex scenarios.
\end{abstract}

\section{Introduction}\label{sec:intro}

\input{sections/intro.tex}

\input{sections/model.tex}

\input{sections/verification.tex}

\section{Conclusions}\label{sec:conclusion}

\input{sections/conclusion.tex}

\bibliographystyle{splncs04}
\bibliography{
	bib/wojtek,
	bib/wojtek-own,
	bib/myrefs
}

\extended{
  \clearpage

  \appendix
  \input{sections/preliminaries.tex}

}%

\end{document}

%% file: sections/intro.tex
In the last 30 years, the world has become densely connected.
\extended{Most IT systems address a complicated network of users, roles, functionalities, and infrastructure elements, often vastly distributed over geographical locations and cultural contexts.}
This results in a considerable space of potential threats, risks, and conflicting interests, that call for systematic (and preferably machine-assisted) analysis. What is more, IT services are implemented by people, with people, and for people. The intensive human involvement makes them hard to analyse beyond the usual computational complexity obstacles.

\para{Voting procedures.}
Voting and elections are prime examples of services that are difficult to specify, hard to verify, and extremely important to the society~\cite{Hao17evoting}. If democracy is to be effective, it is essential to assess and mitigate the threats of fraud, manipulation, and coercion~\cite{meng09critical,Tabatabaei16expressing}. However, formal analysis of voting procedures must consider both the technological side of elections (i.e., protocols, architectures, and implementations) and the human and social context in which it is embedded~\cite{\extended{Basin16humanerrors,}Basin17eve}. The impact of the social factor has become especially evident during the US presidential elections of 2016 and 2020. In 2016, individual voters were targeted before the election by a combination of technology and social engineering to induce emotional reactions that would change their decisions, and possibly swing the outcome of the vote (the Cambridge Analytica scandal). In 2020, a large group of voters was targeted after the election by unfounded claims that severely undermined the public trust in the outcome. In both cases, it is impossible to understand the nature of what happened, and devise mitigation strategies, without the focus on human incentives and capabilities.

\para{Specification and verification of multi-agent systems.}
\emph{Multi-agent systems (MAS)} provide models and methodologies for the analysis of systems that feature interaction of multiple autonomous components, be it humans, robots, and/or software agents. The theoretical foundations of MAS are based on mathematical logic and game theory~\cite{Shoham09MAS\extended{,Weiss99mas,Wooldridge02intromas}}.
In particular, logic-based methods can be useful to formally specify and verify the outcomes of multi-agent interaction~\cite{Dastani10MAS\extended{,Emerson90temporal,Fagin95knowledge,Jamroga15specificationMAS}}.

Formal analysis with multi-agent logics is typically based on model checking~\cite{Baier08mcheck\extended{,Clarke18principles}}. The system is formalized through a network of graphs (or automata) that define its components, their available actions, and the information flow between them. The properties are usually given as \emph{temporal properties}, expressing that a given temporal pattern must (or may) occur, or \emph{strategic properties} capturing the \emph{strategic abilities} of agents and their groups. Especially the latter kind of properties are relevant for MAS; e.g., one may try to capture voter-verifiability as the ability of the voter to verify her vote, and coercion-resistance as the inability of the coercer to influence the behavior of the voter~\cite{Tabatabaei16expressing}.
There are many available model checking tools, though none of them is perfect. Some admit only temporal properties~\cite{Behrmann04uppaal-tutorial\extended{,Dembinski03verics,Kant15ltsmin}}, some focus on the less practical case of perfect information strategies~\cite{\extended{Alur00mocha,Chen13prismgames,}Lomuscio17mcmas}, and the others have limited verification capabilities~\cite{\extended{Akintunde20neuralATL,Kurpiewski19stv-demo,}Kurpiewski21stv-demo}.
Moreover, it is often unclear how to formalize an actual real-life scenario, including the ``right'' model of the system~\cite{Jamroga20Pret-Uppaal} and the formal ``transcription'' of its desirable properties~\cite{Jamroga21anticovid}.

\para{Socio-technical aspects of voting.}
In this paper, we use agent-based methodology to propose and analyze a simple multi-agent model of an actual election, that combines the technological backbone of the voting infrastructure with a model of possible human behaviors. The work is preliminary, in the sense that we do not explore the real breadth of participants' activities that might occur during the vote. Moreover, we mostly look at requirements that can be expressed as trace properties. This is because the computational complexity of the formal analysis turned out prohibitive even for such simple models and properties. We managed to mitigate the complexity by an innovative abstraction technique, but seeing if it scales well enough for realistic models of human interaction remains a subject for future work.

\para{Case study: Polish postal vote of 2020.}
To focus on a concrete scenario, we consider the Polish presidential election of 2020. That was the first time when postal voting was universally available in Poland. Unfortunately, the voting scheme was prepared under pressure, and without the involvement of experts. This might have opened up possibilities for various kinds of ballot fraud, in-house coercion, etc. We propose a preliminary scalable model of the procedure in the form of a Multi-Agent Graph~\cite{Behrmann04uppaal-tutorial,Jamroga22abstraction}, and formalize selected integrity and security properties by formulas of agent logics. Then, we transform the models and formulas so that they can be input to the state-of-art model checker Uppaal~\cite{Behrmann04uppaal-tutorial}, chosen because of its flexible model specification language and user-friendly GUI. 
As expected, the verification of unoptimized models scales rather badly due to the state-space explosion. To improve the performance, we employ a recently developed technique of user-friendly abstraction~\cite{Jamroga22abstraction}, with more promising results.

\extended{
\para{Structure of the paper.}\WJ{revise}
We begin by providing an outline of the Polish postal voting procedure. Then, we present a brief introduction to specification and verification with Uppaal in Section 2.
A MAS Graph for the procedure, together with formal specification of some relevant requirements, are presented in Section 3. This is followed by our experimental results and their discussion in Section X.
XXX? The background terms related to aforementioned abstraction are in the appendix. \WJ{supplementary material?}
}

\extended{
\section{Related Work}

Over the years, the properties of \extended{ \emph{ballot secrecy}, \emph{receipt-freeness}, \emph{coercion resistance}, and \emph{voter-verifiability}}\short{\emph{receipt-freeness} and \emph{coercion resistance}} were recognized as important for an election to work properly.
\extended{In particular, {receipt-freeness} and {coercion-resistance}}\short{They} were studied and formalized in~\cite{Benaloh94receipt,Delaune06coercion,Dreier12formal,Kusters10game,Okamoto98receipt}, see also~\cite{meng09critical,Tabatabaei16expressing} for an overview.

A number of papers used variants of epistemic logic to characterize coercion resistance~\cite{jonker06receipt,kusters09epistemic}.
Moreover, the agent logic \CTLK together with the modeling methodology of {interpreted systems} was used to specify and verify properties of cryptographic protocols, including authentication protocols~\cite{Boureanu16verifSecurity,Lomuscio08securityprots}, and key-establishment protocols~\cite{Boureanu16verifSecurity}.
In particular, \cite{Boureanu16verifSecurity} used variants of the MCMAS model checker to obtain and verify models, automatically
synthesized from high-level protocol description languages such as CAPSL, thus creating a bridge between multi-agent and process-based methods.

\WJ{revise!}

 \cite{killer2019swiss}

human errors etc~\cite{Basin16humanerrors,Basin17eve}
}

\para{Related research.}
Formal verification of voting protocols has been the subject of research for over a decade. Prominent approaches include theorem proving in first-order, linear or higher order logic~\cite{Pattinson15votecounting,Bruni17selene,cortier2018formal,cortier2019beleniosvs,Haines19verified-verifiers,Haines21verifmixnets}, and model checking of temporal, strategic and temporal-epistemic logics~\cite{Jamroga18Selene,Jamroga20Pret-Uppaal,Jamroga21natstrat-voting}.
Most if not all results show that the task is very hard due to the prohibitive computational complexity of the underlying problems.
For example, \cite{cortier2019beleniosvs,cortier2018formal}~conducted a formal analysis of voting protocols using ProVerif, and reported that they had to come up with workarounds for the model in order to the limitations of the  tool.

Modelling and analysis of socio-technical systems is even more difficult because of the vast space of possible human behaviors, and problematic nature of the assumptions usually made about how users choose their actions. The theory of socio-technical systems dates back to the work of Trist and Bamforth in 1940s. In security, perhaps the best studied methodology is based on ceremonies~\cite{Carlos12ceremonies}, in particular the Concertina ceremony~\cite{Bella14concertina,Bella15servicesecurity,Martimiano15ceremony}. Some research has been also based on choreographies~\cite{Bruni21choreographies}.
Moreover, game-theoretic models and analysis have been used in~\cite{Buldas07evoting,Jamroga17preventing,Basin17eve}. 
Here, follow up on the strand based on modeling and verification in multi-agent logics~\cite{Jamroga18Selene,Jamroga20Pret-Uppaal,Jamroga21natstrat-voting}, while trying to put more emphasis on the social part of the system outside of the voting infrastructure.

When analyzing systems that involve human agents, it is important to take into account that they behave differently from the machines, and can make \emph{errors}, or more generally, deviate from the prescribed protocol. This can happen due to a variety of reasons: misunderstanding, inattentiveness, malicious intention, or strategic self-interested action.
Possible deviations from protocol in user behavior have been studied in~\cite{beckert2006method,Basin16humanerrors}, and we follow up on those ideas. To this end, we use the \emph{skilled-human approach} to capture a variety of users' behaviors in our model of postal voting. That is, we extend the protocol specification of an ``honest'' behaviour through a hierarchy of \emph{deviation sets}, i.e., sets of actions that deviate from the protocol and expand the repertoire of the participants.
As pointed out in~\cite{Basin16humanerrors}, there is a trade-off between the breadth of the deviation model and the computational feasibility of the formal analysis.
In fact, our experiments in \autoref{sec:verification} show that even for the skilled human approach only, the explicit state model checking becomes hard enough, and dedicated techniques must be used to mitigate the complexity.

\para{Related verification tools.}
We use the Uppaal model checker~\cite{Behrmann04uppaal-tutorial} in our case study, mainly because of its GUI and a flexible system specification language.
Other verification tools that we considered when preparing the study are:
\begin{itemize}[noitemsep,topsep=2pt,parsep=0pt,partopsep=0pt]
	\item \textit{MCMAS} \cite{lomuscio2017mcmas}: a state-of-art OBDD-based symbolic model checker for agent-based systems. The system is described using ISPL (Interpreted Systems Programming Language), and the requirements are specified as formulae of strategic or temporal-epistemic properties;   
	\item \textit{Tamarin-prover} \cite{meier2013tamarin}: a tool for security protocol verification and not a model checker per se. The system specification language is based on  multiset rewriting theories, and the requirements are specified as first-order temporal properties;
	\item \textit{STV} \cite{kurpiewski2019stv,Kurpiewski21stv-demo}: an experimental toolbox for explicit-state model checking of strategic properties; at the moment, custom input models are not fully supported and may lack documentation; 
	\item \textit{ProVerif} \cite{blanchet2016modeling}: an automated cryptographic protocol verifier, in the symbolic (Dolev-Yao) model. The protocol representation is based on Horn clauses; it can be used for proving secrecy, authentication and equivalence properties. %
\end{itemize}

\noindent Among the above tools, only STV, Uppaal, and (to lesser extent) Tamarin provide a graphical view of the system structure. Of those, only Uppaal allows for \emph{interactive graphical system specification}, which we claim to be crucial in modelling and analysis of voting protocols. Real-life voting procedures include the interaction of numerous participants, each of them with a possibly different agenda and capabilities. Furthermore, the behaviour of most participants is characterized with a mixture of controllable and uncontrollable nondeterminism. In consequence, interactive GUI is crucial if we want to ensure that the model we verify and the one we want to verify are the same thing, cf.~\cite{Jamroga20Pret-Uppaal} for discussion.

Moreover, Uppaal (and, to a smaller extent, STV) allow for parameterized specification of the system, without forcing the designer to program a dedicated model-generator (e.g., as in the verification of SELENE protocol with MCMAS in \cite{Jamroga18Selene}).

%% file: sections/model.tex
\section{Postal Voting Procedure}\label{sec:procedure}

Postal voting is one of the oldest forms for voting.
In its simplest version, it is easy to setup for the authorities and easy to follow for the voters.
On the other hand, it can be susceptible to ballot fraud, lacks verifiability, and opens up potential for vote buying and coercion.
What is more, only basic mechanisms of recovery are possible (e.g., cancelling the whole elections in case of irregularities).
This sometimes leads to controversial ad hoc decisions when dealing with the irregularities~\cite{euronews2021dutch}.

The postal voting procedure employed for the Polish Presidential Election in 2020 is no exception.
There was an overall impression that the procedure had been prepared in haste~\cite{ow2020nieprawidlowosci,mk2020kartki} and with no proper research on the existing postal voting schemes that were proposed and used during the last two decades, such as~\cite{benaloh2013verifiable,killer2019swiss}.
For example, voter authentication is based on the assumption that a voter's national identification number (PESEL) is secret, which is hardly the case in real life.
Moreover, there are various ways how authorities can delete votes, e.g., by sending invalid ballots to districts with anti-government majority~\cite{mk2020kartki,lublin2020kartki,zakopane2020kartki}.
In this paper, we make the first step towards systematic modeling and analysis of this kind of threats.

\begin{figure}[!t]
    \centering
    \includegraphics[width=0.50\textwidth]{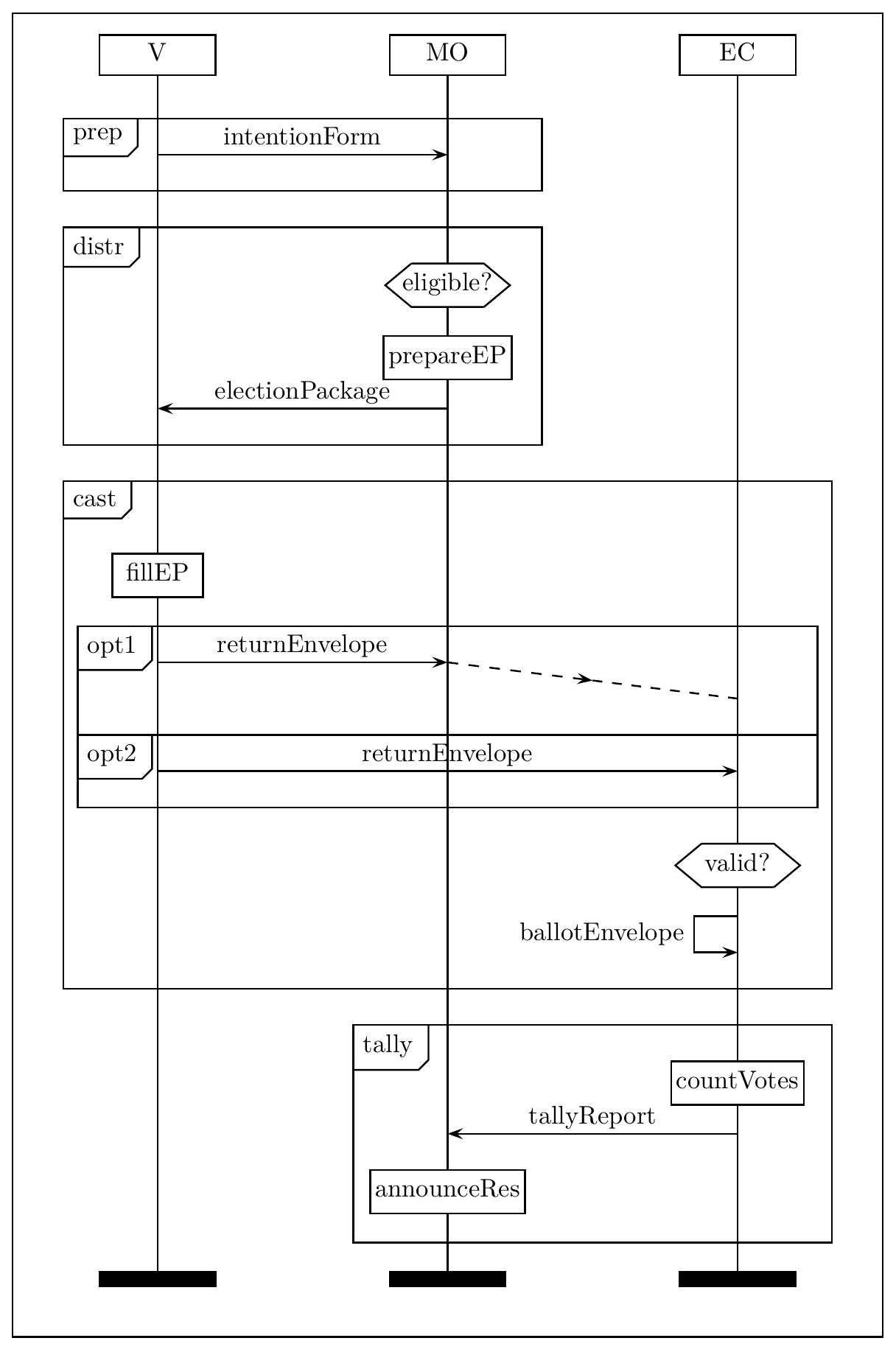}
    \caption{A simplified diagram of the voting process}\label{fig:msc}
    \vspace{-0.55cm}
\end{figure}

\subsection{Postal Voting Procedure for the 2020 Presidential Election}

The rules for organizing the election of the President of Poland, with the possibility of postal vote, were published on June 2\extended{~\cite{act2020june2}} and June 3, 2020\extended{~\cite{regulation2020june3}}; the date of the election was set to June 28, 2020.
A complete list of legal acts defining the election procedure can be found in~\cite{isap2020list,pkw2020june}\extended{, see also~\cite{act2020june8,act2020june10,regulation2020june12,ordinance2020vlist} for additional details}.
For the postal vote, the regulations (mostly concerning the time limits) vary based on the voter's location and her current quarantine status.
We focus on the non-expat and non-quarantined voters,\extended{\footnote{
  Changing this to another type of the voter (or even adding extra ones) would only require  changing values of few configuration variables. However, this would also add unnecessary (in the context of satisfiability of requirements being discussed) complexity to the model. }}
but the protocol for the other types of voters is nearly the same.

The protocol consists of several, partly overlapping phases: the setup which involves expression of intention to vote by post, preparation and distribution of election packages (EPs), casting of the vote, validation of votes\extended{ on the election day}, and tallying, see \autoref{fig:msc}.

\para{Setup.}
A voter expresses her intention to vote by post to its local municipal office (\MO) at the latest 12 days before the day of election (\EDay).
This can be done in either oral, written, or electronic form.
The intent expression must contain the voter's personal information, such as full name, DOB, id number (PESEL), phone number, email, and residential and postal addresses.
If the voter prefers to collect the \EPack in person, this must be specified instead of a postal address.
Additionally, the voter can request to change municipality assigned to her in the voters' register once before the start of the election.\extended{%
\footnote{
  Voters' register contains a list of eligible voters, their full name, date of birth, id number and residence address. The register is maintained by the municipal office, which can also make changes on request.
  To facilitate access of electoral commissions during the election, it is partitioned into voter lists based on the residence address.
  Its main purpose is to enable verification of voters eligibility and to ensure that each voter can vote at most once. }
It is also possible to obtain a voter certificate, which when provided allows to vote in any election commission of any municipality, however obtaining such will lead to a voter being crossed out from the voters' lists, allowing for in-person voting only.} %
{\setlength{\belowcaptionskip}{-15pt}%
\begin{figure}[!t]
	\centering
    \includegraphics[width=0.65\linewidth]{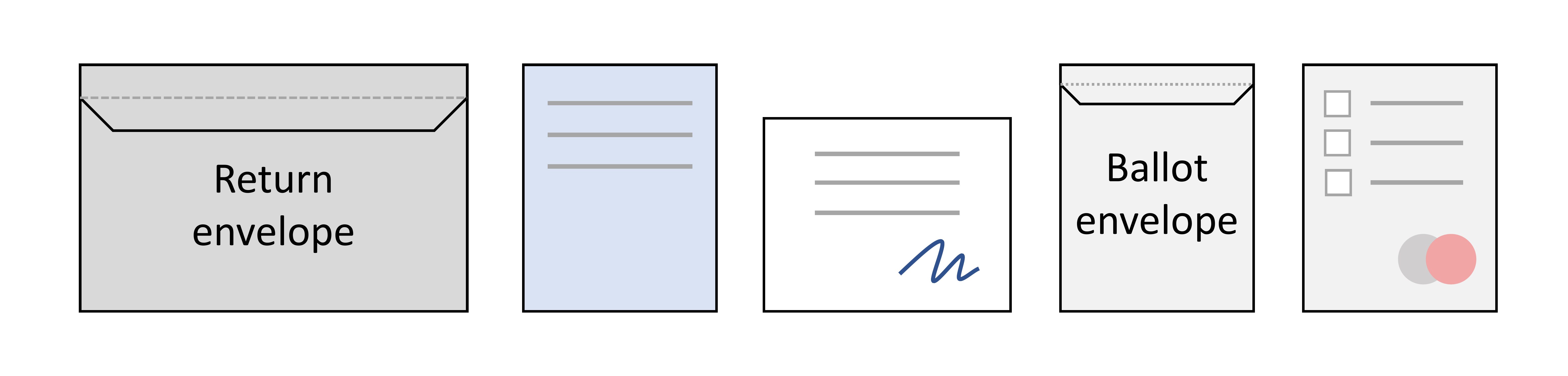}
    \vspace*{-2mm}
	\caption{\EPack content: return envelope, instruction card, voting card, ballot envelope, stamped ballot}\label{fig:epack-content}
\end{figure}}

\para{EP preparation.}
Upon receipt of the intention, a municipal office employee checks the voters' register, and prepares and distributes the \EPack, provided that the applicant is an eligible voter and no required information is missing.
A complete \EPack (\autoref{fig:epack-content}) must contain an instruction, a ballot stamped by both the National Electoral Commission (PKW) and the local electoral commission, a voting card, and two envelopes: one for the ballot and one to be returned.
The \EPacks must be delivered or made available for collection to voters no later than five days before the \EDay.

\para{Casting.}
When a complete \EPack\ is collected, the voter should put a single `X' mark against preferred candidate, put the ballot into a ballot envelope, sign a voting card and place it together with ballot envelope into a return envelope.
Both ballot and return envelope must be sealed.
If there is a deviation in any of the above steps (e.g., the ballot envelope is not sealed), this would invalidate the casting of the vote.
Then, the voter must either send the filled \REnv\ to \MO, where it will be stored until it is passed to the electoral commission (\EC) on the \EDay, or turn it in to the assigned electoral commission.

\para{Validation.}
The \EC\ has to print the voters' list (partitioned according to their municipality) one day before the election at the latest.
This will be used to check the validity of the vote and make sure that no person can vote multiple times.
\extended{In particular, people who requested and obtained a voter's certificate -- a document which allows a voter to go to any election commission for the in-person voting (and not necessarily the one assigned to him) -- are not allowed to vote by post.}
If the voter is eligible and the \REnv is complete, the \BEnv is put into the ballot box.

\para{Tallying.}
At the end of the \EDay, when all \REnvs are collected, the commission opens the \BEnvs and prepares a voting protocol with information on the number of received \REnvs, invalid votes, and the local tally.
The protocol is sent to \MO\ which checks it using a proprietary software. If the errors are within the margins allowed by legislation, the \MO\ accepts the protocol.
Otherwise, the protocol is rejected and the electoral commission must prepare a new one.
When all protocols are accepted and merged, the final tally and the winner are publicly announced.

\extended{
  \begin{figure}[!t]
      \centering
      \fbox{\includegraphics[width=0.9\maxfitwidth]{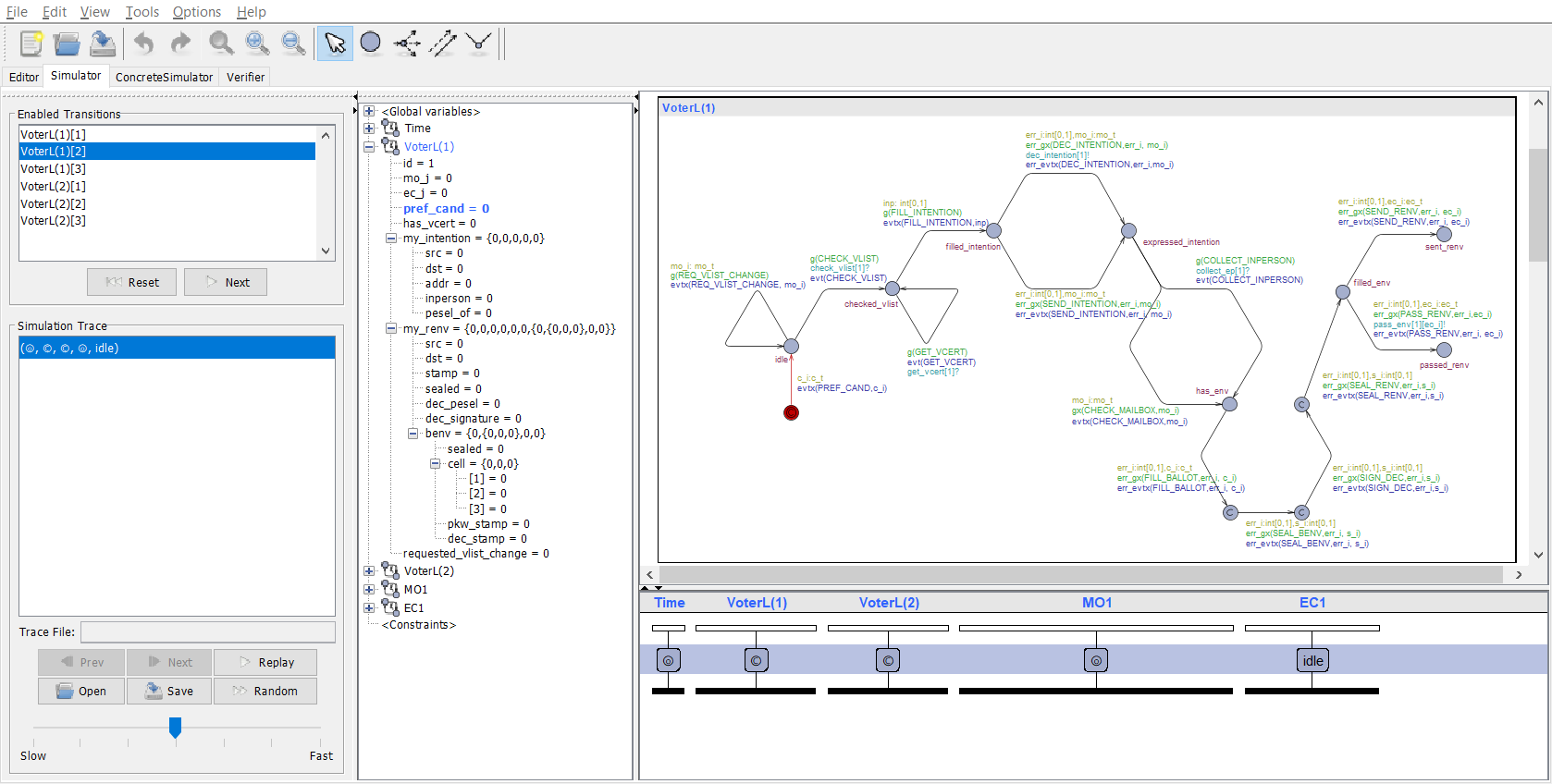}}
      \caption{Uppaal GUI}
  \label{fig:simulator-tab}
  \end{figure}
}%

{\setlength{\belowcaptionskip}{-16pt}%
\begin{figure}[t]
    \hspace{-0.7cm}
    \includegraphics[width=1.1\maxfitwidth]{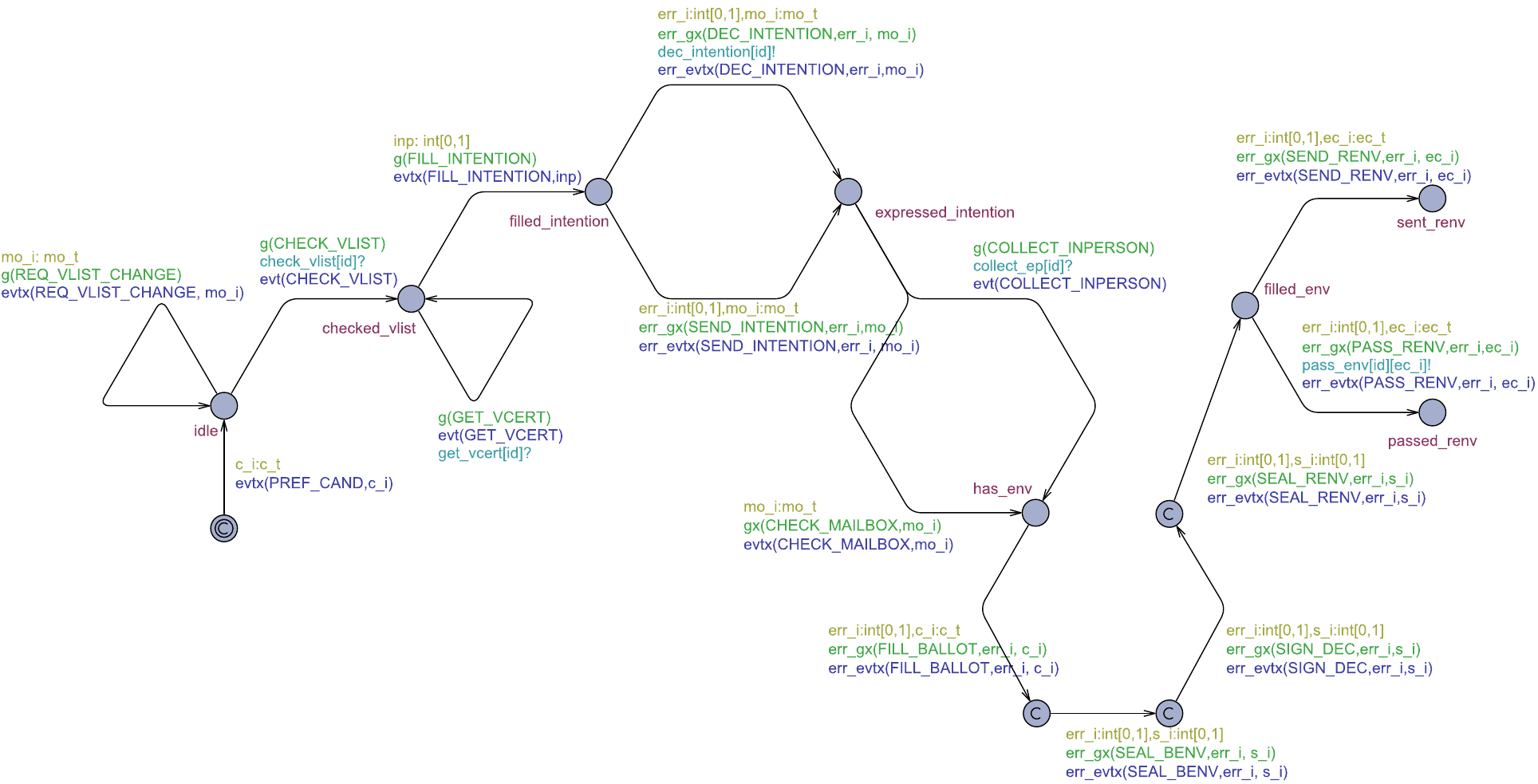}
    \caption{Voter template}
\label{fig:voter-agraph}
\end{figure}}

\section{Formal Model of the Procedure}\label{sec:model}

We can now present our preliminary model of the postal voting protocol.
The code of the model is available at \url{https://github.com/polishpostalvote2020/model}.

\subsection{Automata Networks, MAS Graphs, and Uppaal Model Checker}

We chose the Uppaal model checker as the modeling environment because its user-friendly GUI\extended{ (see \autoref{fig:simulator-tab})} and a flexible system specification language.
The GUI is especially important, as it allows for a preliminary validation of a system specification even at the early stages of modeling.%
\footnote{In Uppaal, system components are defined together with their graph-like (local) representation, which could greatly facilitate in reducing the number of bugs/errors, improve understanding and presentation of the model, and provide more confidence that \textit{we know what we are modelling and if it is actually what wanted to model}.}
A system in Uppaal is represented as a (parameterized) network of (parameterized) finite automata~\cite{Behrmann04uppaal-tutorial}.
The parametrization can be used to define a set of almost identical processes in an easy way.
In order to represent occurrence of events, as well as define the available strategies of participants, we use the extension of automata networks to \emph{MAS graphs}, proposed in~\cite{Jamroga22abstraction}.

A MAS graph allows for the specification of a finite number of agents, possibly interacting via synchronized transitions and/or shared (i.e., global) variables.
Formally, it consists of a set of \emph{shared variables} and a number of \emph{agent templates}, each of the templates being a graph with its own set of \emph{locations}, \emph{edges}, and \emph{local variables}.
An example agent graph is presented in \autoref{fig:voter-agraph}.
The locations are depicted by circles; the initial location (one per agent template) is marked by a double circle.
Committed locations are marked by a circled `C'. Whenever an agent is in such a location, the next transition must involve an edge from the committed location. Those are used to create atomic sequences of events, or to encode synchronization, so that there is no interleaving.
The edges are annotated by selections (yellow), guards (green), synchronizations (teal) and updates (blue). %
A selection is a statement of the form \texttt{var:type}, which binds the identifier to a value from the given range in a non-deterministic way.
These can be used to define a set of parameterized edges in an easy way (both for reading and writing the model). %
Uppaal allows having function calls for the updates and for the guards if it does not lead to side effects (i.e., without modifying value of any variable). %

\extended{
  \begin{figure}[!t]
      \centering
      \includegraphics[width=0.4\maxfitwidth]{images/MunicipalOffice}
      \caption{Municipal Office template}
  \label{fig:mo-agraph}
  \end{figure}%
  \begin{figure}[!t]
      \centering
      \includegraphics[width=0.3\maxfitwidth]{images/Time}
      \caption{Time singleton}
  \label{fig:time-agraph}
  \end{figure}
}%

\subsection{MAS Graph for the Postal Voting Procedure}

The MAS graph for the procedure consists of the following agent templates: Voter (V, depicted in \autoref{fig:voter-agraph}), Electoral Commission (EC, \autoref{fig:ec-agraph}), Municipal Office (MO%
\footnote{Its template consists of a single location with multiple self-loop edges; the figure was omitted due to space constraints.}%
), and Time counter (\short{whose graphs are omitted here due to lack of space}\extended{\autoref{fig:ec-agraph} and \autoref{fig:time-agraph}, respectively}).
The numbers of agent instances are denoted by \NVot, \NMun, and \NElec.
Their values, together with the number of candidates \NCand\ are specified as part of the model configuration.
The sole Time agent is used to impose discrete time constraints on the actions of other agents.

We extend the base model (having infallible agents only) by specifying following mistakes for Voter - may initiate communication and send the forms to a wrong MO or EC, may leave ballot or return envelopes unsealed, misplace the cross mark on the ballot, forget to fill or sign the voting card, attempt to vote by post after obtaining voter's certificate, for Municipal Office (only in some explicitly specified experiments) - may prepare and then distribute ballots without a proper stamp, invalidating those. %
This yields a  hierarchy of the (partially ordered) models, allowing us to study the satisfiability of properties on a finer-grained level, and furthermore get a better understanding of the potential impact of human errors on the system.

The agent templates include a variety of behaviours that can result from human errors as well as purposeful misbehavior.
For example, a Voter may not fill the forms properly, or attempting to communicate with the wrong municipal office or electoral commission.
Furthermore, a Municipal Office can send out an invalid ballot by using a photocopied rather than genuine stamp (which actually happened during the election).
In this version, we do not explicitly model the postal services, and thus omit the possible malicious or erroneous actions of postal clerks, or an adversary.%
\footnote{%
It is worth noting that having a modular representation allows extending the model with other types of agents, including adversary, if needed.}
Instead, we focus on analysis of human interactions and possible effects of their mistakes (as deviation from the expected behaviour).
Similarly, we omit rare events, e.g., those that involve the power of attorney for in-person hand-in of the \REnv.

{\setlength{\belowcaptionskip}{-16pt}%
\begin{figure}[!t]
    \centering
    \includegraphics[width=0.7\maxfitwidth]{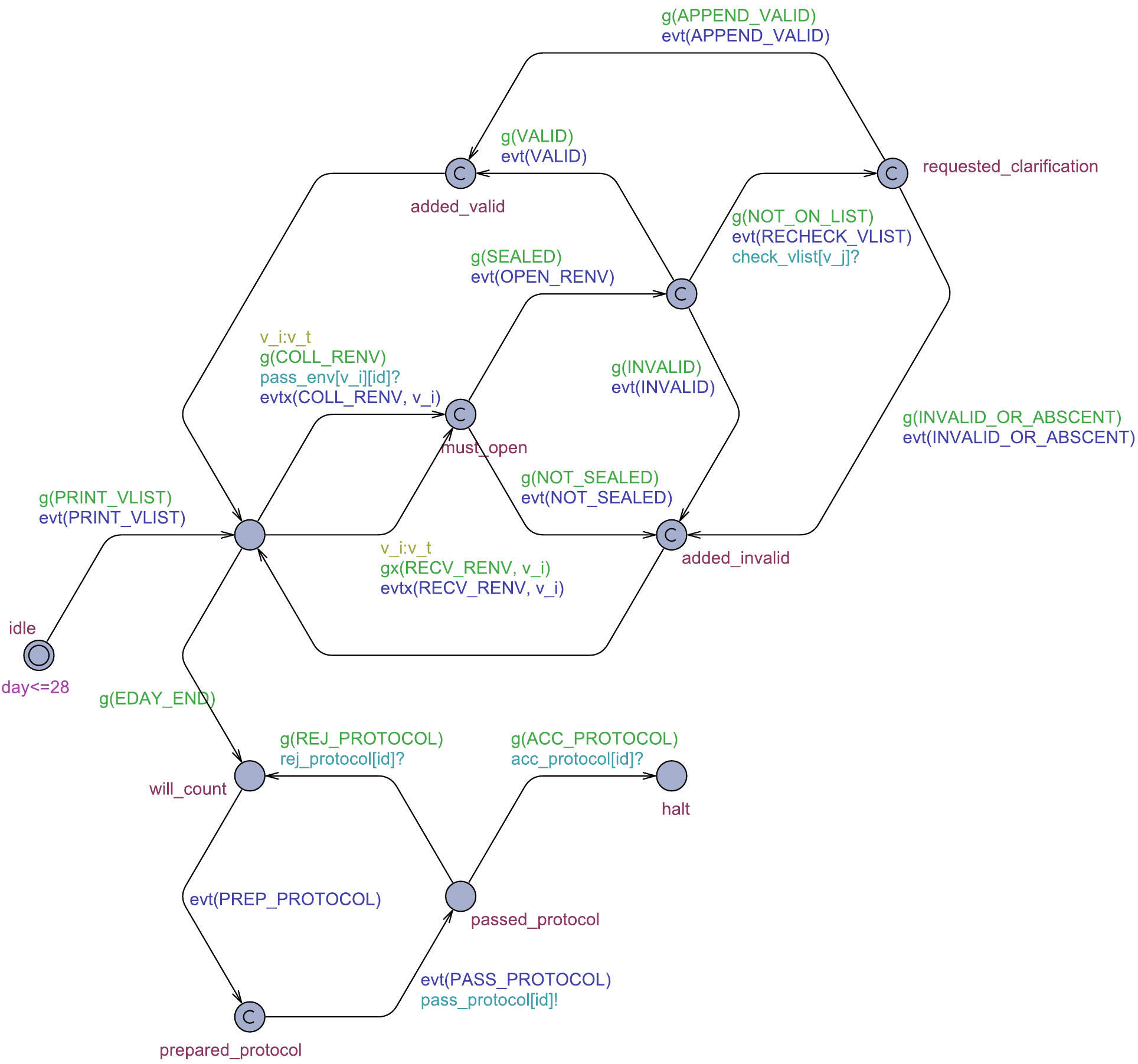}
    \caption{Electoral Commission template}
\label{fig:ec-agraph}
\end{figure}}

The following data structures are used within the templates:
\begin{Verbatim}[tabsize=4,fontsize=\scriptsize,vspace=2pt]
typedef int[1,NC] c_t;              // candidate id
typedef int[1,NV] v_t;              // voter id
typedef int[-NMO,-1] mo_t;          // municipal office id
typedef int[-NMO-NEC,-NMO-1] ec_t;  // election commission id
typedef int[-NMO-NEC,NV] addr_t;    // domain of all agents ids
typedef struct{
	addr_t src;       // sender id
	addr_t dst;       // receiver id
	addr_t addr;      // address for EP delivery
	bool inperson;    // in-person collection preference
	v_tx pesel_of;    // PESEL number
}IntentionForm;
typedef struct{
	bool sealed;        // envelope sealed
	bool pkw_stamp;     // National Electoral Commission stamp
	bool dec_stamp;     // district electoral commission stamp
	int[0,2] cell[c_t]; // number of Xs near candidate cell
                        // (2 for any number greater than 1)
}Benv;                  // ballot envelope
typedef struct{
	addr_t src;          // sender id
	addr_t dst;          // recipient id
	Benv benv;
	mo_tx stamp;         // envelope stamp
	bool sealed;         // envelope was sealed
	bool dec_signature;  // voter's card signed
	v_tx dec_pesel;      // voter's card PESEL
}Renv;                   // return envelope
typedef struct{
	addr_t src;    // sender id
	addr_t dst;    // recipient id
	Renv renv;
}ElectionPackage;  // election package
typedef struct{
	addr_t mo_addr;  // assigned municipality
	addr_t ec_addr;  // assigned commission
	cmt_t comment;   // nominal field for
	bool changed;    // recently changed
}v_record            // an entry in the voters registry
\end{Verbatim}

\if{false}
\begin{itemize}
    \item \texttt{typedef int[1,\NC] c\_t}: a candidate;
    \item \texttt{typedef int[1,\NV] v\_t}: a voter;
    \item \texttt{typedef int[-\NMO,-1] mo\_t}: a municipal office;
    \item \texttt{typedef int [-\NMO-\NEC,\NMO-1] ec\_t}: a election commission;
    \item \texttt{typedef int[-\NMO-\NEC,\NV] addr\_t}: a domain of involved agent identifiers;
    \item %
\begin{Verbatim}[tabsize=4]
typedef struct{
	addr_t src;
	addr_t dst;
	addr_t addr;
	bool inperson;
	v_tx pesel_of;
}IntentionForm;
\end{Verbatim}
a tuple of sender and receiver IDs, postal address for \EPack\ delivery, a boolean in-person collection is preferred and a PESEL number,
    \item %
\begin{Verbatim}[tabsize=4]
typedef struct{
	bool sealed;
	bool pkw_stamp;
	bool dec_stamp;
	int[0,2] cell[c_t];
}Benv;
\end{Verbatim}
ballot envelope, which can either be sealed or not, and contains a ballot that may have required stamps, some number of `X's near each candidate cell (where 2 denotes any number greater than 1);
\item %
\begin{Verbatim}[tabsize=4]
typedef struct{
	addr_t src;
	addr_t dst;
	Benv benv;
	mo_tx stamp;
	bool sealed;
	bool dec_signature;
	v_tx dec_pesel;
}Renv;
\end{Verbatim}
return envelope, which may be sealed, and has ballot envelope and may have sender and receiver IDs, stamp of the municipality which issued it, signed voter's card with voter's PESEL;
\item %
\begin{Verbatim}[tabsize=4]
typedef struct{
	addr_t src;
	addr_t dst;
	Renv renv;
}ElectionPackage;
\end{Verbatim}
election package contains addresses of sender and recipient and a return envelope with blank forms.
    \item %
\begin{Verbatim}[tabsize=4]
typedef struct{
	addr_t mo_addr;
	addr_t ec_addr;
	cmt_t comment;
	bool changed;
}v_record
\end{Verbatim}
    an entry in the voters registry, containing a one's id, assigned \MO\ and \EC, boolean whether the it was already changed during the ongoing elections and a comment nominal field (storing the information if voter certificate was issued, intention was received, \EPack\ prepared, sent or collected, vote received).
\end{itemize}
\fi

Notice that the space of active agent names (or rather their unique identifiers) is partitioned with respect to the agent type.
This allows to model intended parties for a given communication instance in an easy way.%
\extended{\footnote{
  One of implicit assumptions for the model is that every agent can perform only a single role.}}
All the variables in Uppaal are assigned with either explicit initial or implicit default values.
Thus, sometimes aliases which extend the existing data types will be used (e.g., \texttt{c\_tx} for \texttt{int[0,\NC]}).

Additionally, to improve readability and keep the graphical view `clean' we use the following convention for the edge labels:
\begin{itemize}[noitemsep,topsep=0pt,parsep=0pt,partopsep=0pt]
    \item all updates are of the form \texttt{evt(EVT\_CODE)} or \texttt{evtx(EVT\_CODE,EVT\_PARAMS)};
    \item all guards are of the form \texttt{g(EVT\_CODE)} or \texttt{gx(EVT\_CODE,EVT\_PARAMS)}.
\end{itemize}

\noindent Where \texttt{EVT\_CODE} is a string of characters (i.e., named value from the pre-defined enum of events), representing action name by means of a natural language. %
The sets of available/specified actions are disjoint among agents of a distinct type.
On the level of the model code, functions \texttt{evt/evtx} and \texttt{g/gx} are declared using switch statement matching passed argument of \texttt{EVT\_CODE} to its case clause, which should facilitate towards the modularity of a model and higher integrity of its graphical presentation.

%% file: sections/verification.tex
\section{Verification}\label{sec:verification}

\extended{
  ===> To specify requirements Uppaal admits a fragment of \CTLs, excluding the ``next'' and ``until'' modalities.
  Consequently, all supported formulas %
  are of the following types: ``\emph{reachability}'' ($\Epath\DiamondOp{p}$), ``\emph{liveness}'' ($\Apath\DiamondOp{p}$), ``\emph{safety}'' ($\Epath\BoxOp{p}$ and $\Apath\BoxOp{p}$) and an Uppaal specific ``\emph{leads-to}'' ($\Apath\BoxOp{(p \Then\Apath\DiamondOp{p})}$).
  Its semantics allows for non-maximal paths, in this paper we will usually use $\Epath\DiamondOp{p}$ and $\Apath\BoxOp{p}$ for specifying the desired properties.

  As it was already discussed in \cite{jamroga2020towards} limitations in Uppaal's requirement specification lead to it's verification capabilities being rather limited.
  Interestingly, in the next section we will see that some of the critical requirement can still be represented by means of supported formulas, despite those limitations.
  <===
}

As the next step, we specify some relevant properties of the voting system by formulas of multi-agent logics, in particular the branching-time temporal logic \CTL and the strategic logic \ATL.
Then, we transform them to a form that can be interpreted by Uppaal, and run the model checking.
To mitigate the impact of state-space explosion, we use abstraction on variables, proposed in~\cite{Jamroga22abstraction}.

\subsection{Specification of Properties}\label{sec:properties}

Following~\cite{Tabatabaei16expressing,Jamroga18Selene,Jamroga20Pret-Uppaal,Jamroga21natstrat-voting}, we use the formulas of multi-agent logics to specify some  interesting requirements on the election system.
In particular, we use the branching-time temporal logic \CTL~\cite{Emerson90temporal} and strategic logic \ATL~\cite{Alur02ATL}.
We focus on the following requirements\footnote{%
Authors would want to stress out that the given properties are just an example, merely for illustration purpose, and they should not be viewed as complete list of requirements. 
}:
\begin{description}[noitemsep,topsep=2pt,parsep=0pt,partopsep=0pt]
    \item[$(\textit{\bstuff})$] The number of correctly received ballots cannot exceed the number of sent ballots (a weak variant of resistance against ballot stuffing);
    \item[$(\textit{\valvote})$] For every voter, cast vote must be properly recorded and reflected by the tally (\tallypropname);
    \item[$(\textit{\moblock})$] The authorities should have no strategy to invalidate certain votes, even when the voters' preferences are known (no strategic ballot removal).
    \extended{see here: https://www.tandfonline.com/doi/abs/10.1080/13510347.2019.1574755?journalCode=fdem20}
\end{description}

We formalize $(\textit{\bstuff})$ and $(\textit{\valvote})$ by the following $\CTL$ formulas:
{%
\setdisplayskips[2pt]
\begin{align*}
    \varphi_{\textit{\bstuff}} & \equiv  \Apath\BoxOp ({\textstyle\sum_{i=1}^{\NB}b\_received_i} \leq {\textstyle\sum_{j=1}^{\NV}}(ep\_sent_j))
    \\
    \varphi_{\textit{\valvote}}     & \equiv  \Apath{\BoxOp(\textit{elec}\_\textit{end}\wedge \textit{voted}_{i,j} \Rightarrow tallied_{i,j})}
\end{align*}
}

\noindent where $\NB$ and $\NV$ stand for the maximum number of ballots and voters accordingly, $i\in\NV$ is an arbitrary fixed voter, and $j\in \NC$ a candidate. The $(\textit{\bstuff})$ says that, for all possible execution paths ($\Apath$) and all future time points ($\BoxOp$), the sum of received ballots must not exceed the number of sent election packets. 
Similarly, $({\textit{\valvote}})$ says that if election is closed and voter has cast her vote for candidate $j$, then it must be tallied for $j$.

Furthermore, we formalize $(\textit{\moblock})$ by the following formula of \ATL (in fact, we formalize the negation of $(\textit{\moblock})$ and focus on ballot removal by a Municipal Office, thus expressing that the MO can strategically remove ballots):
{\setdisplayskips[2pt]%
\begin{align*}
    \varphi_{\neg\textit{\moblock}} & \equiv  \coop{\textit{MO}_k}{\BoxOp\bigwedge_{i\in\NV} (\textit{vreg}_{i,k} \wedge \textit{pref}_{i,j} \wedge \textit{elec}\_\textit{end} \Rightarrow  \neg \textit{tallied}_{i,j})}
\end{align*}}

\noindent where $i\in\NV$, $j\in \NC$ and $k\in\MO$ is a municipal office.
The reading of $\varphi_{\neg\textit{\moblock}}$ is: ``Municipal office $k$ has a strategy ($\coop{\textit{MO}_k}$) such that, no matter what the other agents do, at all future time points ($\BoxOp$)
for any voter $i$ if registered in this municipality ($\textit{vreg}_{i,k}$), prefers candidate $j$ ($\textit{pref}_{i,j}$) and election is already closed ($\textit{elec}\_\textit{end}$), then her vote will not be tallied correctly ($\neg\textit{tallied}_{i,j}$).

\subsection{From Agent Logics to Uppaal Specifications}\label{sec:properties-uppaal}

Our formalization of resistance to ballot stuffing and tally integrity has a straightforward transcription in Uppaal specification language:
\begin{description}[noitemsep,topsep=2pt,parsep=0pt,partopsep=0pt]
    \item [($\varphi_{\textit{\bstuff}}$)] \texttt{A[] (b\_recv<=ep\_sent) },
    \item [($\varphi_{\textit{\valvote}}$)]
    \texttt{A[] (Time.end and (Voter(i).sent\_renv or Voter(i).passed\_renv) \\ imply recorded\_link[i]==Voter(i).pref\_cand) },
\end{description}
where \texttt{b\_recv}, \texttt{ep\_sent} and \texttt{recorded\_link} are auxiliary variables added only for the verification of related property, and representing the number of received ballots, the number of sent election packages, and the mapping from voters to the way their votes have been tallied.
Moreover, \texttt{pref\_cand} is the voter's local variable storing her preferred candidate, \texttt{sent\_renv} and \texttt{passed\_renv} are labels of the corresponding locations in the voter's agent graph (see \autoref{fig:voter-agraph}).

Unfortunately, Uppaal does not offer the verification of strategic abilities thus does not admit \ATL operators.
To deal with that, we propose to approximate formula $\varphi_{\neg\textit{\moblock}}$ by its \emph{under-approximation} $\varphi^{-}_{\neg\textit{\moblock}}$ and \emph{over-approximation} $\varphi^{+}_{\neg\textit{\moblock}}$, both of which are \CTL formulas that satisfy the following conditions:
{\setdisplayskips[2pt]%
\begin{align*}
     & M\models_{\CTL} \varphi^{-}_{\neg\textit{\moblock}} &  & \Rightarrow &  & M\models_{\ATL} \varphi_{\neg\textit{\moblock}}      \\
     & M\models_{\ATL} \varphi_{\neg\textit{\moblock}}      &  & \Rightarrow &  & M\models_{\CTL} \varphi^{+}_{\neg\textit{\moblock}}
\end{align*}}

\noindent That is, whenever $\varphi^{-}_{\neg\textit{\moblock}}$ is true in a model, $\varphi_{\neg\textit{\moblock}}$ must also be true there.
Moreover, if $\varphi^{+}_{\neg\textit{\moblock}}$ is false in a model, $\varphi_{\neg\textit{\moblock}}$ must also be false.
We use the following approximations:
{\setdisplayskips[0pt]%
\begin{eqnarray*}
    \varphi^-_{\neg\textit{\moblock}} & \equiv & \Apath{\BoxOp\bigwedge_{i\in\NV} (\textit{vreg}_{i,k} \wedge \textit{pref}_{i,j} \wedge \textit{elec}\_\textit{end} \Rightarrow  \neg \textit{tallied}_{i,j})} \\
    \varphi^+_{\neg\textit{\moblock}} & \equiv & \Epath{\BoxOp\bigwedge_{i\in\NV} (\textit{vreg}_{i,k} \wedge \textit{pref}_{i,j} \wedge \textit{elec}\_\textit{end} \Rightarrow  \neg \textit{tallied}_{i,j})}
\end{eqnarray*}%
}
This follows the intuition that, if $\psi$ is guaranteed to always hold on all execution paths ($\Apath\BoxOp\psi$), then it must also hold when MO plays strategically ($\coop{MO}\BoxOp\psi$).
Moreover, if MO has a strategy to maintain $\psi$ ($\coop{MO}\BoxOp\psi$), then $\psi$ must always hold on at least one path ($\Epath\BoxOp\psi$).

Now, formula  $\varphi^-_{\neg\textit{\moblock}}$ can be fed directly to Uppaal.
Unfortunately, this is not the case for the upper approximation $\varphi^+_{\neg\textit{\moblock}}$, as Uppaal does not interpret the $\Epath\BoxOp$ combination of \CTL operators correctly.\footnote{%
    The satisfaction of \CTL operators in a transition system is interpreted over \emph{maximal runs}, i.e., ones that are either infinite or end in a state with no outgoing transitions.
    In contrast, Uppaal looks at \emph{all finite runs}. While this does not change the semantics of $\Apath\BoxOp$ and $\Epath\DiamondOp$, the interpretation of both $\Apath\DiamondOp$ and $\Epath\BoxOp$ becomes nonstandard. }
On the other hand, Uppaal's \texttt{E[]} combination is an over-approximation of the \CTLs $\Epath\BoxOp$ combination.
Thus, we can use it to provide ``over-over-approximation'' of the original specification, which finally obtains the following list of Uppaal inputs:
\begin{small}
\begin{description}[noitemsep,topsep=0pt,parsep=0pt,partopsep=0pt]
    \item [($\varphi^-_{\neg\textit{\moblock}}$)] \texttt{A[] forall(i:v\_t)(Time.end and vlist[i].mo\_addr==k and vpref[i]==j imply recorded\_link[i]!=j) },
    \item [($\varphi^{++}_{\neg\textit{\moblock}}$)] \texttt{E[] forall(i:v\_t)(Time.end and vlist[i].mo\_addr==k and vpref[i]==j imply recorded\_link[i]!=j) },
\end{description}
\end{small}
where \texttt{vlist} refers to the voters' registry.
{\setlength{\belowcaptionskip}{-19pt}%
\begin{figure}[t]
\begin{minipage}[t]{0.45\textwidth}
\begin{Verbatim}[%
    frame=topline,
    framesep=3mm,
    tabsize=4,
    fontsize=\scriptsize, 
    vspace=0pt,
    label=Concrete model
]
typedef struct{
    bool sealed;
    bool pkw_stamp;
    bool dec_stamp;
    int[0,2] cell[c_t];
}Benv;
typedef struct{
    addr_t src;
    addr_t dst;
    Benv benv;
    mo_tx stamp;
    bool sealed;
    bool dec_signature;
    v_tx dec_pesel;
}Renv;
typedef struct{
    addr_t src;
    addr_t dst;
    Renv renv;
}ElectionPackage;
\end{Verbatim}
\end{minipage}\hfill
\begin{minipage}[t]{0.45\textwidth}
\begin{Verbatim}[frame=topline,framesep=3mm,tabsize=4,fontsize=\scriptsize, label=Abstact model]
typedef struct{
    bool invalid;
    c_tx cell;
}Benv;
typedef struct{
    addr_t dst;
    bool invalid;
    Benv benv;
}Renv;
typedef struct{
    bool sent;
    Renv renv;
}ElectionPackage;
\end{Verbatim}
\end{minipage}%
\caption{Fragment of code resulting from the abstraction, where all evaluations invalidating a vote are merged into a single variable (for \REnv\ and \BEnv)}
\label{fig:renv-abstr}
\end{figure}}

\subsection{Mitigating State Space Explosion by Abstraction of Variables}

One of the biggest challenges in the practical application of model checkers is the so called \emph{state-space explosion}.
Typically, model checking involves the generation of a huge state/transition graph that includes all the possible states (i.e., configurations) of the system. Clearly, the number of such configurations is exponential in the number of processes and their components -- in our case, the number of agent instances and their local variables~\cite{clarke2018handbook}.
This is easy to see in our experimental results for formula $\varphi_{\textit{\bstuff}}$, see \autoref{tab:bstuff}(left), the part under ``$\varphi_{\textit{\bstuff}}$ (concrete),'' with the clear exponential growth of the verification time $t$ and memory use $\virt$.
As a consequence, the verification of $\varphi_{\textit{\bstuff}}$ on the model presented in \autoref{sec:model} scales up to only 2 voters, 1 municipal office, and 3 electoral commissions for an election with 3 candidates.

Mitigating state-space explosion has been an important topic of research for over 30 years.
The most important techniques include partial-order reduction~\cite{Peled93representatives\extended{,godefroid1996partial,Gerth99por,Jamroga20POR-JAIR}},
symbolic verification~\cite{McMillan93symbolic-mcheck}, bounded and unbounded model checking~\cite{McMillan02unbounded\extended{,Penczek03ctlk,Lomuscio07tempoepist}} and state/action abstraction~\cite{\extended{Cousot77abstraction,}Clarke94abstraction\extended{,Godefroid02abstraction}}.
In particular, abstraction is an intuitive model reduction method, based on the idea of clustering ``similar'' states of the system (so called \emph{concrete states}) into \emph{abstract states}, hopefully reducing the model to a manageable size.
The actual clustering must be carefully crafted. On the one hand, it must only remove information that is irrelevant for the verification of a given property, otherwise the verification results for the abstract model will be inconclusive with respect to the original model (so called ``concrete model'').
On the other hand, it has to remove sufficiently much of the concrete model, so that the model checking becomes efficient.

In this work, we use an intuitive and easy to use abstraction scheme, based on the removal of variables from agent graphs~\cite{Jamroga22abstraction}. %
The method allows to select a subset of local variables to be removed (possibly with a subset of locations to serve as the scope of the abstraction).
For example, the name of the voter's preferred candidate is irrelevant for the verification of resistance to ballot stuffing, hence the corresponding variables can be omitted in the voters' agent graphs.
The abstraction generates \emph{two} abstract models.
The first one \emph{under-approximates} the concrete model, in the sense that if formula $\psi$ returns \texttt{true} on the abstract model, it must be also be true in the concrete model.
The second \emph{over-approximates} the concrete model, i.e., if $\psi$ returns \texttt{false} on the abstract model, it must be also be false in the concrete model.

Alternatively, the user can define a mapping from the variables to a fresh variable that merge some of the information that used to be stored in the removed variables.
For example, we might map the complex representation of all potential ballot faults (unsealed ballot envelope, missing stamps, more than one `X' on the ballot, for one or more candidates) to a single boolean variable \texttt{invalid}, see \autoref{fig:renv-abstr}.

In general, these abstraction parameters should be picked firstly to reduce the number of induced states of global model, and secondly to match the property, so that verification is conclusive. 
Naturally, there are variety of ways to chose fitting parameters for a given property; it is also possible that two distinct properties are matched by the same ones, and therefore same abstract models.
We refer to~\cite{Jamroga22abstraction} for the formal definitions, correctness proofs, and extensive discussion of the method.

\subsection{Verification Experiments}

Based on the input prepared in the previous sections, we have conducted a number of model checking experiments.
All the results presented here were obtained with Uppaal 4.1.24 (32 bit) on a laptop with Intel i7-8665U 2.11 GHz CPU, running Ubuntu 20.04 on WSL2 with 4 GB RAM.
The outcome of the experiments is shown in Tables~\ref{tab:bstuff} and~\ref{tab:moblock}. The notation is as follows:
\begin{itemize}[noitemsep,topsep=2pt,parsep=0pt,partopsep=0pt]
\item
  \emph{conf} denotes the configuration of the experiment, i.e., the number of voters, Municipal Offices, Electoral Commissions, and election candidates;
\item
  \emph{Sat} reports the verification output, i.e., whether the model checker returned \texttt{true} or \texttt{false};
\item
  \emph{t} and \emph{\virt} show the time and memory used in the verification, with \texttt{memout} indicating that the model checking process ran out of memory.
\extended{Note that the included time measurements could have been affected by other processes running on OS.}
\end{itemize}

\autoref{tab:bstuff}(left) presents the experimental results for our formalization of weak resistance to ballot stuffing.
The formula has turned out to be true in all the completed verification runs. However, as already observed, the verification scales rather badly due to state-space explosion.
To mitigate that, we reduced the models by abstracting away the identity of candidates, and simplified the data structures representing the intention form and the election package.
Moreover, we mapped variables \texttt{b\_recv} and \texttt{ep\_sent} to a single fresh variable \texttt{ballot\_diff = ep\_sent-b\_recv}, so that the requirement specification became \texttt{AG(ballot\_diff>=0)}.
The results for model checking the under-approximating abstract model are also presented in \autoref{tab:bstuff}(left). %

Since the output of under-approximation was \texttt{true}, we conclude that $\varphi_{\textit{\bstuff}}$ is also true in the original (concrete) model.
Note that, for some configurations, the abstraction allowed to run the verification faster by orders of magnitude. 
Moreover, it allowed for the model checking of scenarios with 3 voters, 1 MO, 1 EC, and 3 candidates, i.e., one more voter than in the concrete case. This might seem slight, but in some cases 3 voters are necessary to demonstrate non-trivial attacks on a voting system~\cite{Arapinis16threeVoters}.

For \tallypropname ($\varphi_{\textit{\valvote}}$), we used  formula $\varphi_{\textit{\valvote}}$ proposed in Section~\ref{sec:properties-uppaal}.
In this case, we additionally generated the under-approximating abstract models obtained by (a) mapping all the candidate names except $j$ to a fresh value $j'$, (b) for all voters other than $i$, removing their memory of the choices associated with their intention forms and election packages after they send those.
The results in \autoref{tab:bstuff}(right) show that the verification output was not conclusive, i.e., they do not imply whether $\varphi_{\textit{\valvote}}$ is true or false in the original model.
However, the verification becomes conclusive under the assumption that voter $i$ makes no errors and strictly follows the protocol, see the rightmost part of the table.
In that case, we get \texttt{true} as the output, thus concluding the original property holds as well.

Lastly, the experimental results for strategic ballot removal ($\varphi_{\neg\textit{\moblock}}$) are presented in \autoref{tab:moblock}.
The table first shows the (inconclusive) output of model checking for under-approximation w.r.t. the formula, under-approximation w.r.t. the formula and the model, and over-approximation w.r.t. the formula. Thus, the original property might (but does not have to) be satisfied.
Then, we fix a strategy for MO in the model, so that the municipal office sends a ballot with invalid stamps whenever the voter intends to vote for the ``unwelcome'' candidate.\extended{\footnote{
  A situation underlying this scenario actually took place for some of the voters abroad \cite{mk2020kartki}, there were also reports of this problem by voters from Lublin \cite{lublin2020kartki} and Zakopane \cite{zakopane2020kartki}. }}
The results in the rightmost part of \autoref{tab:moblock} show now conclusive output: the under-approximation is true, so the original property must be true as well.
Thus, the proposed strategy indeed achieves the goal specified by $\varphi_{\neg\textit{\moblock}}$.
Note that it is essential to couple matching formula- and model-related approximations, otherwise the procedure is not sound.
In our case, this meant using the \emph{under-approximating} formula $\varphi^-_{\neg\textit{\moblock}}$ with the \emph{under-approximating} abstract model of the procedure.

\begin{table}[!p]
\centering
\caption{Experimental results for model checking of $\varphi_{\textit{\bstuff}}$ (left) and $\varphi_{\textit{\valvote}}$ (right)}
	\smallskip
    \scalebox{0.65}{
    	\input{images/tables/table1.tex}
	}\quad
    \scalebox{0.65}{
    	\input{images/tables/table1-rightside.tex}
	}
    \label{tab:bstuff}
\end{table}

\begin{table}[!p]
\centering
\caption{%
	Experimental results for model checking of $\varphi^-_{\neg\textit{\moblock}}$ and $\varphi^{++}_{\neg\textit{\moblock}}$ in a model, where \MO\ may prepare \EPack\ without a valid stamp }
    \smallskip
	\scalebox{0.65}{
    	\input{images/tables/table4.tex}
	}
    \label{tab:moblock}
\end{table}

Despite the technical limitations on the number of voters, it was possible to discover and verify attacks violating $\varphi_{\valvote}$ and $\varphi_{\moblock}$ respectively. 
For the next step (which remains as a subject of future work), we would want to scale up the model to larger or even unbounded number of voters (the latter is currently beyond technical feasibility of the tool), or to come up with  rigorous arguments that certain number of voters is sufficient for certain cases.

%% file: images/tables/table1.tex
\begin{tabular}[t]{ |c||cHHrrH|cHHrrHHHHHHH|}
    \hline
    \multirow{2}{*}{conf}        & \multicolumn{6}{c|}{$\varphi_{\textit{\bstuff}}$ (concrete)} & \multicolumn{6}{c}{$\varphi_{\textit{\bstuff}}$ (abstract)} & \multicolumn{6}{H|}{Abstract2}                                                                                                                                                                                                    \\
    \cline{2-19}
        & Sat                           & \#St$_s$                      & \#St$_e$                       & t (s) & \virt ({\textsc{mb}}) & res ({\textsc{mb}}) & Sat                     & \#St$_s$ & \#St$_e$ & t (s) & \virt ({\textsc{mb}}) & res ({\textsc{mb}}) & Sat                     & \#St$_s$ & \#St$_e$ & t (s) & \virt ({\textsc{mb}}) & res ({\textsc{mb}}) \\
    \hline
    1,1,1,1 & \small{\texttt{true}}         & 2079                          & 3221                           & 0.07   & 31        & 9        & \small{\texttt{true}}   & 0        & 1463     & 0.07   & 30        & 9        & \small{\texttt{true}}   & 0        & 1463     & 0.052   & 30        & 9        \\
    1,1,1,2 & \small{\texttt{true}}         & 4157                          & 6441                           & 0.10   & 31        & 9        & \small{\texttt{true}}   & 0        & 1463     & 0.07   & 30        & 9        & \small{\texttt{true}}   & 0        & 1463     & 0.05   & 30        & 9        \\
    1,1,1,3 & \small{\texttt{true}}         & 6235                          & 9661                           & 0.12    & 31        & 10       & \small{\texttt{true}}   & 0        & 1463     & 0.07   & 30        & 9        & \small{\texttt{true}}   & 0        & 1463     & 0.05   & 30        & 9        \\
\hline
\end{tabular}%

%% file: images/tables/table1-rightside.tex
\begin{tabular}[t]{ |c||cHHrrH|cHHrrH||cHHrrH|}
    \hline
    \multirow{2}{*}{conf} & \multicolumn{6}{c|}{$\varphi_{\textit{\valvote}}$  (concrete)} & \multicolumn{6}{c||}{$\varphi_{\textit{\valvote}}$ (abstract)} & \multicolumn{6}{c|}{$\varphi_{\textit{\valvote}}$ (honest abstract)}                                                                                                                                                                                                                                                \\
    \cline{2-19}
                          & Sat                                                          & \#St$_s$                                                     & \#St$_e$                                                            & t (s)   & \virt ({\textsc{mb}}) & res (MB) & Sat                    & \#St$_s$              & \#St$_e$ & t (s)   & \virt ({\textsc{mb}}) & res (MB) & Sat                   & \#St$_s$ & \#St$_e$ & t (s) & \virt ({\textsc{mb}}) & res (MB) \\
    \hline
    1,1,1,1               & \small{\texttt{false}}                                       & 1853                                                         & 2645                                                                & 0.07   & 31                    & 9        & \small{\texttt{false}} & 0                     & 1333     & 0.06   & 30                    & 9        & \small{\texttt{true}} & 0        & 745      & 0.05 & 30                    & 9        \\
    1,1,1,2               & \small{\texttt{false}}                                       & 3704                                                         & 5287                                                                & 0.09   & 31                    & 9        & \small{\texttt{false}} & 0                     & 2664     & 0.06   & 31                    & 9        & \small{\texttt{true}} & 0        & 1489     & 0.06 & 30                    & 9        \\
\hline
\end{tabular}%

%% file: images/tables/table4.tex
\begin{tabular}{ |c||cHHrrH|cHHrrH|cHHrrH||cHHrrH|}
    \hline
    \multirow{2}{*}{conf} & \multicolumn{6}{c|}{$\varphi^{-}_{\neg\textit{\moblock}}$ (concrete)} & \multicolumn{6}{c|}{$\varphi^{-}_{\neg\textit{\moblock}}$ (abstract)} & \multicolumn{6}{c||}{$\varphi^{++}_{\neg\textit{\moblock}}$ (concrete)} & \multicolumn{6}{c|}{$\varphi^{-}_{\neg\textit{\moblock}}$ (str. abstract)}                                                                                                                                                                                                                                                                                                                                                                                                                                                   \\
    \cline{2-25}
                          & Sat                                                               & \#St$_s$                                                           & \#St$_e$                                                           & t (s)                                                                   & \virt ({\textsc{mb}}) & res ({\textsc{mb}}) & Sat                                           & \#St$_s$              & \#St$_e$                                     & t (s) & \virt ({\textsc{mb}}) & res ({\textsc{mb}}) & Sat                   & \#St$_s$              & \#St$_e$ & t (s)   & \virt ({\textsc{mb}}) & res ({\textsc{mb}}) & Sat                   & \#St$_s$ & \#St$_e$ & t (s) & \virt ({\textsc{mb}}) & res ({\textsc{mb}}) \\
\hline
\end{tabular}%

%% file: sections/conclusion.tex
In this paper, we demonstrate how multi-agent methodology can be used to specify and analyze the impact of human aspects on security and integrity of voting protocols. 
We also argue that postal voting protocols provide good material for case studies, that will hopefully increase our understanding of the subject, and help to design better protocols.

Speaking in more concrete terms, we propose a preliminary analysis of the Polish postal vote used in the presidential election of 2020.
We use Multi-Agent Graphs to represent the participants and their interaction, and formulas of multi-agent logics \CTL and \ATL to encode interesting properties.
Then, we transform those to match the input of the state-of-art model checker Uppaal.
This way, the obtained models are given an intuitive visual representation and a modular structure that allows for easier modifications and detection of errors.
We also use a recently proposed method of state abstraction by variable removal to reduce the models and mitigate state-space explosion.
The method is guaranteed to preserve the truth of universal \CTL specifications and thus generates correct-by-design abstract models. No less importantly, it is easy to use, requires almost no technical knowledge from the user, and provides significant savings in terms of the verification performance.

Despite the limitations of Uppaal in the expressive power of its property specification language, we have managed to conclusively verify the selected properties for nontrivial configurations of voters, electoral commissions, and candidates.
To this end, we used approximation over formulas (by providing weaker or stronger versions of the original requirements) and approximation w.r.t. models (by generating appropriate abstract models).
Choosing the right approximations was by no means obvious and required some skill.
We believe, however, that this is inevitable: successful formal analysis of real-life scenarios requires both science and art. With more than a little bit of understanding and domain knowledge.

For the future work we plan to employ alternative verification tool to conduct analysis of a broader scope of interesting properties, and adapt our abstraction methods for that. 
One of the interesting directions is to adopt a methodology defining a families of possible human mistakes as in \cite{sempreboni2020x}, where human agent deviates from the protocol through a combination of skipping, modifying, or adding action(s), or in \cite{bella2022modelling}, which also advocates using the epistemic modal logic distinguishing between knowledge and possession.

\para{Acknowledgements.}
The work was supported by NCBR Poland and FNR Luxembourg under the PolLux/FNR-CORE project STV (POLLUX-VII/1/2019).

%% file: sections/preliminaries.tex
\section{MAS Graphs and Variable Abstraction}\label{sec:masgraphs}

\YK{TODO: add some parts on abstraction from the other paper?}

\begin{definition}\label{def:agent-graph}
    An \emph{\agname}\extended{, describing the behaviour of an agent,}	is a tuple $\agsym=(\textit{Var}, \textit{Loc}, l_0, \textit{Cond}, g_0, \textit{Act}, \textit{Effect}, \hookrightarrow)$ consisting of:
    \begin{itemize}[noitemsep,parsep=0pt]
        \item $\Var$: a finite set of typed \emph{variables}. The set of possible valuations of the variables is denoted by $\Eval(\Var)$, and given by the Cartesian product of their domains;
        \item $\Loc$: a finite set of \emph{locations} (nodes in the graph);
        \item $l_0\in\Loc$: the initial location;
        \item $\Cond$: a set of logical conditions over $\Var$ (also called \emph{guards}), possibly using arithmetic operators;
        \item $g_0\in \Cond$: the initial condition\extended{, required to be satisfied at the initial location};
        \item $\Act$: a set of \emph{actions}, with $\tau\in\Act$ standing for the ``do nothing'' action;
        \item $\textit{Effect}: \Act\times \Eval(\Var) \mapsto \Eval(\Var)$: the \emph{effect} of actions on the values of variables;
        \item $\hookrightarrow\,\subseteq\Loc\times\Cond \times\Chan \times\Act \times\Loc$: the set of labelled edges defining the local \emph{transition relation}. An edge with synchronization $ch\neq -$ can only be taken synchronously with an edge with $\overline{ch}$ in another agent graph.
    We will often write $l\xhookrightarrow{g:ch\,\alpha}l'$ instead of $(l,g,ch,\alpha,l')\in\,\hookrightarrow$, and omit $ch=-$ from the labelling.
    \end{itemize}
\end{definition}

\begin{definition}\label{mas-graph}
    A \emph{\magname}\ is a multiset of \agnames additionally parameterized by a set of shared (global) variables.
    We assume w.l.o.g.~that all variables have unique names (e.g., by prefixing each name of a local variable by the name of its agent graph).
\end{definition}

\begin{definition}
    Let $\magsym = \multiset{\Var_{sh},\agsym_1,\ldots,\agsym_n}$ be a
    \magname having a set of shared variables $\Var_{sh}$.
    The \emph{combined \magname} of $\magsym$ is the agent graph\\
    $\agsym_{\magsym} = (\Var, \Loc, l_0, \Cond, g_0, \Act, \Effect,\hookrightarrow)$, where:

    \begin{itemize}[noitemsep,topsep=0pt,parsep=0pt,partopsep=0pt]
        \item $\Var = \bigcup_{i=1}^{n}\Var_i$,
        \item $\Loc = \Loc_1\times\ldots\times\Loc_n$,
        \item $l_0 = (l_{0,0}, \ldots, l_{n,0})$,
        \item $\Cond = \bigcup_{i=1}^{n}\Cond_i$,
        \item $g_0 = g_{0,0}\wedge\ldots\wedge g_{n,0}$,
        \item $\Act = \bigcup_{i=1}^{n}\Act_i$.
    \end{itemize}

    \noindent Relation $\hookrightarrow$ is obtained inductively by the following rules (where $l_i,l'_i\in\Loc_i$, $l_j,l'_j\in\Loc_j$, $c\in \ChanId_i\cap\ChanId_j$
    for two	\agnames $\agsym_i$ and $\agsym_j$ of distinct indices	$1 \leq i,j\leq n$):

    \begin{small}
    \begin{tabular}{c@{\qquad}c}
    $\begin{prooftree}
            l_i \lhook\joinrel\xrightarrow{g_i:c!\alpha_i}_i l_i' \wedge l_j \lhook\joinrel\xrightarrow{g_j:c?\alpha_j}_j l_j'
            \justifies
            (l_i,l_j)\lhook\joinrel\xrightarrow{g_i\wedge g_j: (\alpha_j\updcomp \alpha_i) }(l_i',l_j')
            \end{prooftree}$	
    &
    $\begin{prooftree}
        l_i \lhook\joinrel\xrightarrow{g_i:\alpha_i}_i l_i'
        \justifies
        (l_i,l_j)\lhook\joinrel\xrightarrow{g_i: \alpha_i }(l_i',l_j)
        \end{prooftree}$
    \\
    \\
    $\begin{prooftree}
            l_i \lhook\joinrel\xrightarrow{g_i:c?\alpha_i}_i l_i' \wedge l_j \lhook\joinrel\xrightarrow{g_j:c!\alpha_j}_j l_j'
            \justifies
            (l_i,l_j)\lhook\joinrel\xrightarrow{g_i\wedge g_j: (\alpha_i\updcomp \alpha_j) }(l_i',l_j')
            \end{prooftree}$
    &
    $\begin{prooftree}
        l_j \lhook\joinrel\xrightarrow{g_j:\alpha_j}_j l_j'
        \justifies
        (l_i,l_j)\lhook\joinrel\xrightarrow{g_j: \alpha_j }(l_i,l_j')
        \end{prooftree}$
    \end{tabular}
    \end{small}

    \noindent Lastly, the effect function is defined by: %
    $$\small\textit{Effect}(\alpha,\eta) =
    \begin{cases}
        \Effect_i(\alpha,\eta) &\text{if } \alpha\in \Act_i\\
        \Effect(\alpha_i,\Effect(\alpha_j,\eta)) &\text{if } \alpha = \alpha_i\updcomp\alpha_j
    \end{cases}$$
\end{definition}

\begin{definition}
    A \emph{model} is a tuple $M=(\textit{St}, I, \rightarrow, AP, L)$, where
    $\textit{St}$ is a set of states,
    $I\subseteq \textit{St}$ is a non-empty set of initial states,
    $\longrightarrow  \subseteq \textit{St}\times \textit{St}$ is a transition relation,
    $AP$ is a set of atomic propositions,
    $L:\textit{St}\to 2^{AP}$ is a labelling function.
    We assume $\longrightarrow$ to be serial, i.e., there is at least one outgoing transition at every state.
\end{definition}

\begin{definition}\label{def:unwrapping}
    The \emph{\unwrapping} of an \agname\ $\agsym%
    \extended{=(\textit{Var}, \textit{Loc}, l_0, \textit{Cond}, g_0, \hookrightarrow, \textit{Act}, \textit{Effect})}$ %
    is a model $\mathcal{M}(\agsym) = (\textit{St}, I, \longrightarrow, AP, L)$, where:
    \begin{itemize}
        \item $\textit{St} = \Loc \times \Eval(\Var)$,
        \item $I = \{ \abracket{l_0, \eta} \in \textit{St}\mid \eta\in \textit{SAT}(g_0) \}$,
        \item $\longrightarrow\ = \longrightarrow_0\ \cup \{(s,s)\in \textit{St} \times \textit{St} \mid \lnot\exists s'\in \textit{St}\dott s\longrightarrow_0 s'\}$, where
            $\longrightarrow_0\ =\{(\abracket{l,\eta},\abracket{l',\eta'})\in \textit{St} \times \textit{St} \mid \exists\ l\xhookrightarrow{g:\alpha}l': \eta\in \textit{SAT}(g) \wedge \eta'=\Effect(\alpha, \eta)\}$.
            Note that $\longrightarrow$ adds loops wherever necessary to make the relation serial;
        \item $AP = \Loc \cup \Cond$,
        \item %
        $L(\abracket{l,\eta})=\{l\}\cup\{g\in \Cond \mid \eta \in \textit{SAT}(g)\}$.
    \end{itemize}
    Nodes and edges in an agent graph correspond to \emph{sets} of states and transitions, defined by its unwrapping.\\
    The unwrapping $\unwrap(\magsym)$ of a \magname $\magsym$ is given by the unwrapping of its combined graph.
\end{definition}

\begin{definition}\label{def:path}
    Let $M$ be a model.
    A \emph{run} in $M$ is a sequence of states $s_0s_1\dots$, such that $s_i\longrightarrow s_{i+1}$ for every $i$.
    {A \emph{path} is an infinite run.}
    The sets of all runs in $M$, all paths in $M$, and all paths starting from state $s$ are denoted by
    $\textit{Runs}(M)$, $\textit{Paths}(M)$, and $\textit{Paths}(s)$.
    In addition, $\textit{Runs}^{t}(M)$ denotes the set of all runs in $M$ of length $t\in\mathbb{N}^+\cup\{\infty\}$.
\end{definition}